# Magnetic Anomalies in a New Manganocuprate

# $Gd_3Ba_2Mn_2Cu_2O_{12}$

**Sudhindra Rayaprol[1,*], Kiran Singh[2], S. Narayana Jammalamadaka[3], Niharika Mohapatra[3], N. K. Gaur[2] and E. V. Sampathkumaran[3]**

[1]*UGC-DAE Consortium for Scientific Research, Mumbai Centre, B.A.R.C, Mumbai – 400085, India*

[2]*Department of Physics, Barkatullah University, Bhopal – 462026, India*

[3]*Tata Institute of Fundamental Research, Homi Bhabha Road, Colaba, Mumbai – 400005, India*

[*]Corresponding Author:

Dr. Sudhindra Rayaprol

UGC-DAE Consortium for Scientific Research

Mumbai Centre

R-5 Shed, BARC, Trombay

Mumbai – 400 085

INDIA

Tel:    +91-22-25594727 / 25505327

Fax:    +91-22-25505402

Email: rayaprol@gmail.com / sudhindra@csr.ernet.in

**Abstract**

The manganocuprate compound $Gd_3Ba_2Mn_2Cu_2O_{12}$ (Gd-3222) has been synthesized by conventional solid state reaction method and its magnetic behavior has been studied by *dc* and *ac* magnetization (M) and heat capacity (C) measurements as a function of temperature (T). This compound crystallizes in a tetragonal structure (space group *I4/mmm*). We find that this compound exhibits three magnetic transitions, around 2.5, 4.8 and 9 K, as inferred from *dc* and *ac* magnetic susceptibility ($\chi$) data. However, no evidence for a well-defined $\lambda$-anomaly is found in C(T) above 1.8 K, though there is a gradual upturn below about 10 K. An application of a magnetic field results in a peak around 5K, while ac $\chi$ appears to show a very weak frequency dependence below 9 K. Isothermal M curve at 1.8 K exhibits a weak hysteresis without any evidence for saturation even at fields as high as 120 kOe. These results imply that this compound undergoes a spin-glass-like freezing at low temperatures, though the exact nature of the magnetic transition at 10 K is not clear. The magnitude of the magnetocaloric effect, as inferred from M and C data, is quite large over a wide temperature range below 50 K peaking around 4 K.

**Keywords**:    $Gd_3Ba_2Mn_2Cu_2O_{12}$, Manganocuprate, Anomalous magnetism

**PACS:**    **75.50.-y, 75.30.Sg**

# 1. INTRODUCTION

The research on perovskite compounds during past two decades has been activated following the discovery of high temperature superconductivity and colossal magnetoresistance phenomena. The investigations on the crystal chemistry of copper and manganese based compounds have also picked up momentum [1-4]. Some recent reports have shown that it is possible to synthesize a new inter-grown perovskite structure containing both Cu and Mn [5-8]. The structure of the resultant compound is a layered one and is expressed as $Ln_3Ba_2Mn_2Cu_2O_{12}$ ($Ln$-3222), which is an intergrowth of "$K_2NiF_4$" and "$R123$" type structure, i.e., an intergrowth of rocksalt type and perovskite type structure [5]. Due to the ionic size effects only the compounds $Ln$ = Sm, Eu and Gd are known to exist in the $Ln$-3222 series [6]. The work of Field $et$ $al$ [7] shows that it is very difficult to synthesize $Sm_3Ba_2Mn_2Cu_2O_{12}$ (Sm-3222) in single phase form [7]. Matsubara $et$ $al$ [8] has carried out magnetic and transport studies on $Eu_3Ba_2Mn_2Cu_2O_{12}$ (Eu-3222). However, to our knowledge, no detailed study on the analogous Gd compound has been reported till to date except for some preliminary structural studies by Matsubara $et$ $al$ [9]. In this article, we report the results of our detailed magnetic investigations on this compound. The results reveal that the magnetism of this compound is quite complex at low temperatures (<10 K), pointing towards spin-glass-like freezing below 4.8 K. In addition, the magnetocaloric effect (MCE) as inferred from magnetic entropy is large over a wide temperature range (below 50 K) at low temperatures, peaking in the vicinity of 4 K.

# 2. EXPERIMENTAL DETAILS

The polycrystalline sample of the compound, $Gd_3Ba_2Mn_2Cu_2O_{12}$ has been synthesized by the conventional solid state reaction method. High purity (all with stated purities better than 99.9%) starting materials viz., $Gd_2O_3$, $BaCO_3$, $MnO_2$ and CuO were taken in stoichiometric ratios and grounded thoroughly in an agate mortar under acetone. The samples were calcined at 1000°C for 4 hrs in powder form. The calcined powders were compacted into 12mm diameter pellets by applying the hydraulic pressure of 8 tons and sintered in ambient condition at 1100°C for 30 hours with one intermittent grinding. The $Gd_3Ba_2Mn_2Cu_2O_{12}$ samples were annealed in oxygen atmosphere at 850°C for 6 hrs to obtain better phase purity. Phase formation was confirmed by the X-ray diffraction (Cu $K_\alpha$ radiation). In addition, back-scattered image obtained with scanning electron microscope confirms homogeneity of the specimen and energy dispersive x-ray (EDX) analysis confirms the composition.

Magnetic measurements were carried out employing commercial magnetometers - a Vibrating Sample Magnetometer, VSM, (Oxford, UK) and as well as a Superconducting Quantum Interference Device, SQUID (Quantum Design, USA). Heat-capacity (C) measurements were carried by relaxation method on a Quantum Design - Physical Property Measurements System (PPMS).

## 3. RESULTS AND DISCUSSIONS

In Fig. 1 we show the Rietveld fitted x-ray diffraction pattern for Gd-3222. Using the structural parameters of Eu-3222 given by Field *et al* [7] as the starting model, the observed XRD pattern for Gd-3222 could be fitted into the tetragonal structure, space group *I4/mmm*. The refinement was carried out using the Fullprof suite program [10].

Fig. 2 shows the *dc* magnetic susceptibility ($\chi$) for Gd-3222 measured in an applied field of 0.5 Tesla. With decreasing temperature, $\chi$ increases gradually with decreasing temperature obeying Curie-Weiss law down to about 30 K. Below 10 K, there is a sudden rise in $\chi$ as if there is a magnetic transition. Interestingly, there is a broad peak in $\chi$(T) around 10 to 20 K, the origin of which is not clear; similar observation was made [11] for single crystalline $ErPd_2Si_2$ (around 8 to 20 K) and it was related to short range magnetic correlations. As the temperature is lowered further, there is a peak around 4 K, followed by a downturn around 2 K. These results imply the existence of three magnetic transitions for this compound. In the data taken in a low field (0.01T) (see figure 3), the 10K-transition is not clearly highlighted, but there is an upturn in $\chi$ below 5K due to the onset of another magnetic transition. The zero-field-cooled and field-cooled curves tend to bifurcate below 5 K; in addition, the broad peak around 10-20K is visible. It may be noted that, from the linear region in the plot of inverse $\chi$ versus temperature (figure 2), the values of paramagnetic Curie temperature ($\theta_p$) and effective Bohr magneton number ($\mu_{eff}$) were obtained: about 13 K and 15.34 $\mu_B$ per formula unit respectively. The values of $\theta_p$ in Gd-3222 is enhanced with respect to that in Eu-3222 [5, 6] in which case the corresponding value was reported to be about 1 K. This may imply that the presence of Gd ions facilitates exchange interaction. The positive sign indicates dominant ferromagnetic exchange. It was noted in Ref. 8 that the value of $\mu_{eff}$ for the Eu analogue falls in the range 8.5-9.4 $\mu_B$ and the enhanced value for the present compound is nearly accountable if one considers theoretical moment from trivalent Gd ions. However, it is rather difficult for us to separate out contributions from Cu and Mn or to infer whether both these ions contribute to magnetism.

The magnetic transitions seen in *dc* $\chi$ measurements are seen in ac $\chi$ data as well. In Fig. 4, ac $\chi(T)$ measured at various frequencies is shown. In the real part, there are peaks at 4.5 and 2.8 K with weak frequency dependence typical of spin-glasses. In addition, frequency dependence of values is observed near 10 K as well, as though all these transitions in some form are glassy in their character. Though the imaginary part is quite noisy, the features due to magnetic transitions are visible.

To throw more light on the magnetism of this compound, we have measured isothermal *dc* magnetization at various temperatures (figure 5). The M(H) curve at 1.8 K curve exhibits a curvature increasing gradually with magnetic field (H) non-linearly, but does not saturate up to field of 12 Tesla. This behavior rules out ferromagnetism. On the other hand, there is a weak hysteresis at low fields. These findings could be consistently interpreted in terms of spin-glass-like freezing at this temperature, as inferred above. As the temperature is raised, say, beyond 4 K, the features remain qualitatively the same, except that the low-field hysteresis is not visible. Even at 50K, M(H) varies non-linearly with H, indicating that short range magnetic correlations well above magnetic transition temperature. At 100 K, M varies linearly with H as the system is in the paramagnetic state.

Further confirmation for the glassy nature of the magnetic transition comes from the heat capacity measurements. In Fig. 6 we show the data as C vs. T as well as C/T vs. T. For H = 0 and 1 T, C(T) does not show any peak above 1.8, thus ruling out the possibility of any long-range antiferromagnetic or ferromagnetic ordering. There is an upturn below about 9 K – the same temperature at which the χ(T) curves reveal a transition. As field is increased to 6 T, a broad plateau like feature starts appearing around 5 K, which shifts to higher temperatures on increasing the field to 10 T. The appearance of this broad feature on the application of magnetic field can be attributed to possible gradual alignment of spins.

Isothermal remanent magnetization ($M_{IRM}$) behavior as a function of time (t) was tracked after cooling the specimen in zero field, then switching on a filed of 0.5 T for 5 mins, and then switched off. While the values fell to a negligibly small value at 7 K, there is a slow decay (stretched exponential behavior) at 1.8 and 3.3 K as a function of time. The observed behavior of $M_{IRM}$ as a function of time at 1.8 and 3.3

K respectively is shown in Fig. 7 along with their fitting to the stretched exponential. This finding supports our conclusion that at these temperatures, the compound behaves a spin-glass.

Finally, we have also inferred the magnetocaloric effect behavior – a topic of great current interest from the applications point of view [12, 13] – as measured by entropy change, $\Delta S$, from the knowledge of isothermal M behavior measured in close intervals of temperature and the C(T) data shown in figure 6. Though we have taken M(H) at several temperatures (every 1 K below 10 K, every 2 K in the range 10- 20 K, every 5 K in the range 20-50 K) , we show the curves for few selected temperatures in figure 5. The well-known Maxwell's relationship was employed to derive $\Delta S$ from magnetization. The results thus obtained are shown in figure 8. It is clear that the values are quite large over a wide temperature range and it peaks around 4 K with value as high as 27 J/mol K for a change in the magnetic field from zero to 10 T. In volume units, it turns out to be about 220 mJ/ccK, comparable to $ErAl_2$ [12, 13]. In figure 8, we have also plotted for two fields the results obtained from the C(T) data presented above in support of the trends observed from M(H) data. While doing so, we adjusted the $\Delta S$ values at 50 K with those obtained from the M(H) data to take into account the entropy change below 2 K.

## 4. CONCLUSIONS

We have studied the magnetic behavior of a new manganocuprate compound $Gd_3Ba_2Mn_2Cu_2O_{12}$. Interestingly, this compound exhibits multiple magnetic transitions and features at low temperatures, unlike the situation in Eu-based analogue as reported in the literature. It is however difficult for us to firmly assign the magnetic transitions (to Gd, Mn and/or Cu?). It is however apparent that the magnetic features are glassy in character, and not strictly ferromagnetic and antiferromagnetic. In addition, this compound exhibits large magnetocaloric effect at low temperatures.

*Acknowledgements*

We thank Kartik K. Iyer for his help in performing experiments. KS duly acknowledges the financial grant from CSIR, New Delhi (India) in the form of Senior Research Fellowship (SRF).

**REFERENCES**

[1]     R. J. Cava, J. Am. Ceram. Soc. 83 (2000) 5.

[2]     J. B. A. A. Elemans, B. van Laar, K. R. van der Veen, and B. O. Loopstra, J. Solid State Chem. 3 (1971) 238.

[3]     A. Gormezano and M. T. Weller, J. Mater. Chem. 7 (1993) 771.

[4]     P. D. Battle, M. A. Green, N. S. Laskey, J. E. Millburn, M. J. Rosseinsky, S. P. Sullivan, and J. F. Vente, Chem. Commun. 6 (1996) 767.

[5]     M. Hervieu, C. Michel, R. Genuouel, A. Maignan and B. Raveau, J. Solid State Chem. 115 (1995) 1.

[6]     I. Matsubara, R. Funahashi, N. Kida, K. Ueno and H. Ishikawa, Physica C 282-287 (1997) 945.

[7]     M. A. L. Field, C. S. Knee, and M. T. Weller, J. Solid State Chem. 167 (2002) 237.

[8]     Ichiro Matsubara, Noriaki Kida, Ryoji Funahashi, Kazuo Ueno, Hiroshi Ishikawa and Nobuhito Ohno, J. Solid State Chem. 141 (1998) 546.

[9]     Ichiro Matsubara, Ryoji Funahashi, Noriaki Kida and Kazuo Ueno, Chem. Lett. (1996) 971.

[10]    J. Rodrigues-Carvajal, Physica B 192 (1993) 55.

[11]  E.V. Sampathkumaran, Niharika Mohapatra, Kartik K Iyer, C.D. Cao, W. Löser, and G. Behr, J. Magn. Magn. Mater. (In press).

[12]  See, a review, K.A. Gschneidner Jr, V.K. Pecharsky, and A.O. Tsokol, Rep. Prog. Phys. 68 (2005) 1479.

[13]  A.M. Tishin, J. Magn. Magn. Mater. 316 (2007) 351.

**<u>Figure Captions</u>**

**Fig.1**  (Color online) Rietveld fitted X-ray diffraction pattern for Gd-3222 compound. The Rietveld refinement parameters obtained from the fitting are $R_p$ = 8.54, $R_{wp}$ = 12.4, $R_{exp}$ = 7.66, $\chi^2$ = 2.61. The refined values of unit cell constants are a = b = 3.8803(4) Å; c = 35.2268 (3) Å.

**Fig. 2**  The magnetic susceptibility $\chi$ (= M/H) measured in a field of 0.5 Tesla. The inverse susceptibility ($\chi^{-1}$) is also shown in the figure. The inset shows the $\chi(T)$ data on an expanded scale below 20 K. The arrows in the inset highlight the observed magnetic transitions at low temperatures.

**Fig. 3**  (Color online) Magnetic Susceptibility measured in field of 0.01 T after zero-field-cooling and field-cooling.

**Fig. 4**  (Color online) Real ($\chi'$) and imaginary ($\chi''$) parts of ac susceptibility for Gd-3222 measured in an ac field of 1 Oe. The lines through the data points serve as guides to the eyes. Arrows in the figure indicates the observed magnetic transitions.

**Fig. 5**  (Color online) Isothermal magnetization for Gd-3222 measured at several temperatures after zero field cooling in each case. The inset shows M-H data for T = 1.8, 4 and 6 K respectively. The curve well below 0.5 T is hysteretic at 1.8K, whereas it is reversible at higher temperatures.

**Fig. 6** (Color online) Specific heat measured at H = 0, 1, 6 and 10 Tesla (T) fields. The bottom panel shows C/T vs. T (temperature). The lines through the data points serve as guides to the eyes.

**Fig. 7** (Color online) Isothermal magnetization ($M_{IRM}$) as a function of time (in minutes) measured at T = 1.8 and 3.3 K. The observed decay in M at these two temperatures could be fitted to the stretched exponential of the form, $M_{IRM}/M_{IRM(t=0)}$ = {P1 + P2exp[-(x/P3)$^{P4}$]}, where P1 – P4 are fitting parameters.

**Fig.8** (Color online) The entropy change as a function of temperature for a variation of the magnetic field, obtained from the magnetization data employing Maxwell's equation. The lines through the data points serve as guides to the eyes. Corresponding data obtained from heat-capacity (C) are also plotted for selected fields.

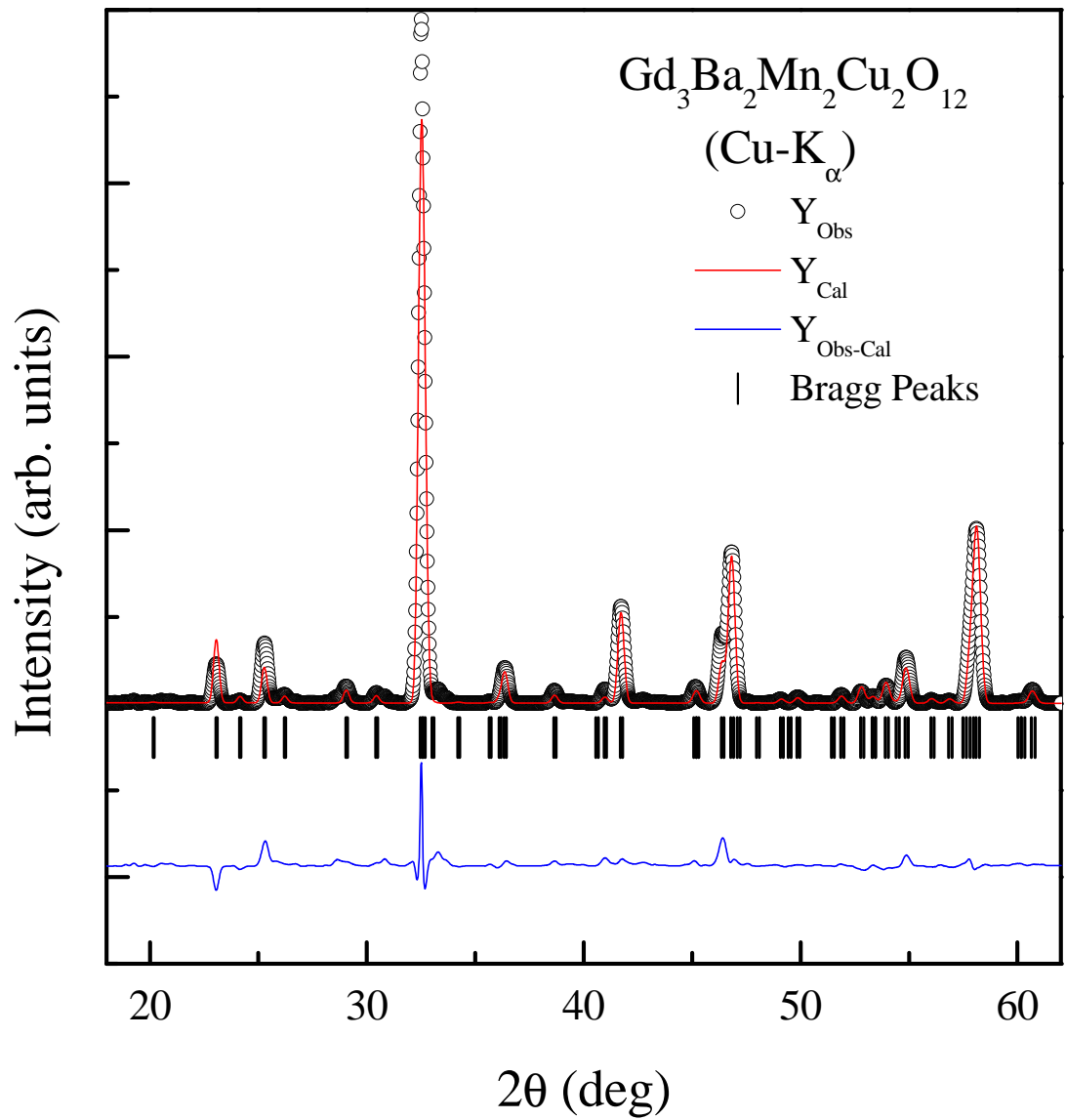

Fig.1 Rayaprol et al

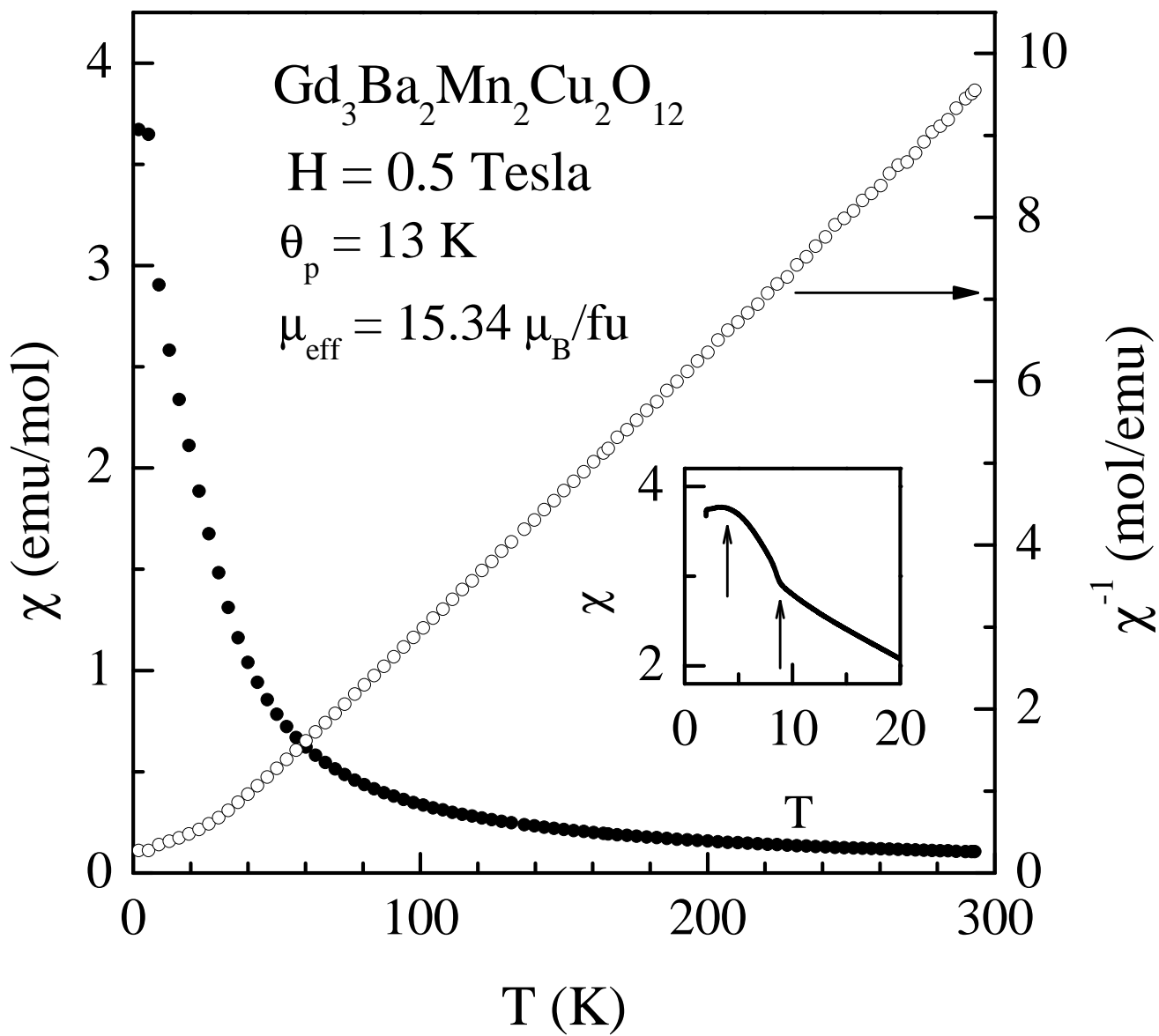

Fig.2. Rayaprol et al

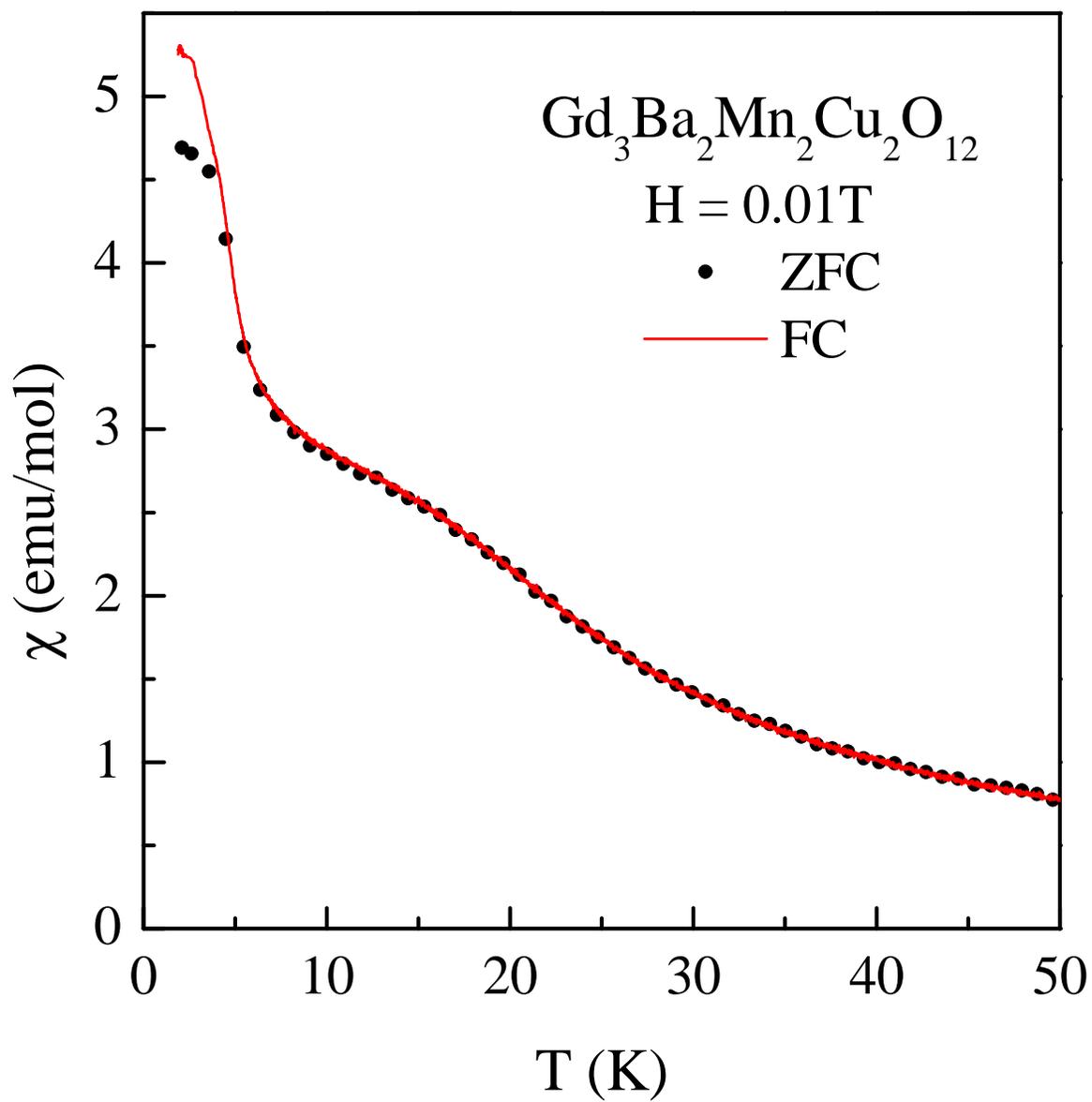

Fig.3 Rayaprol et al

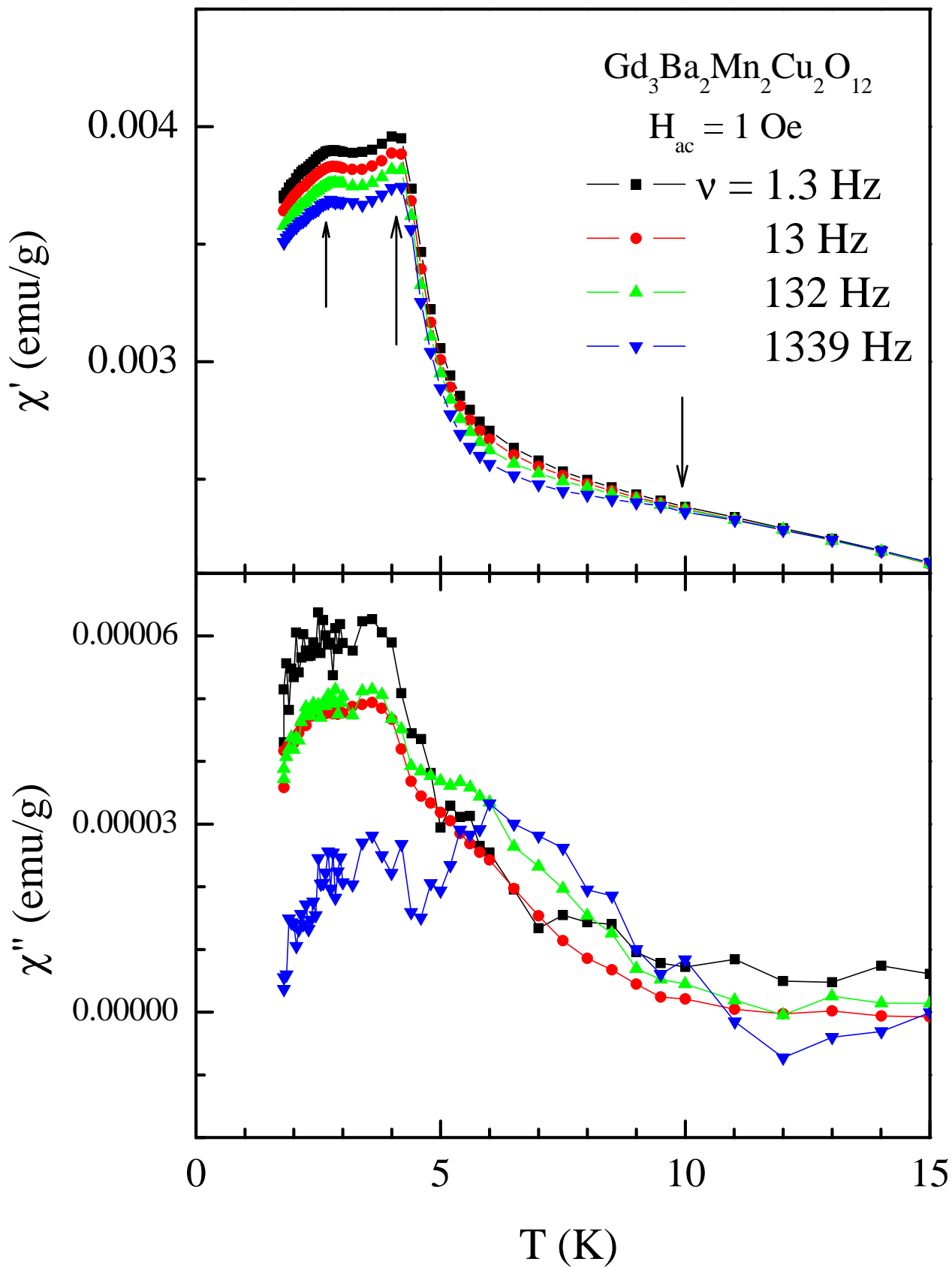

Fig. 4 Rayaprol et al

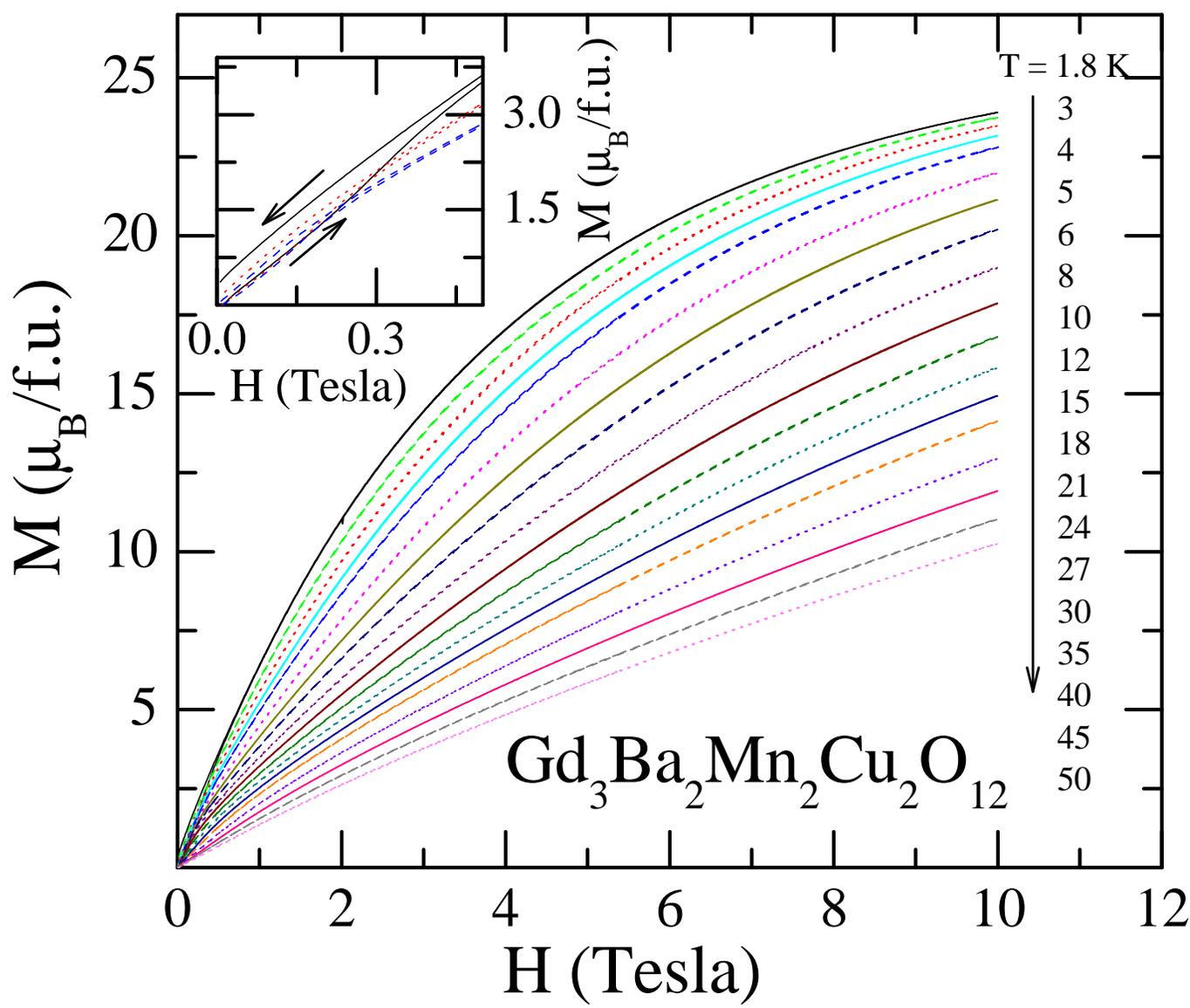

Fig.5 Rayaprol et al

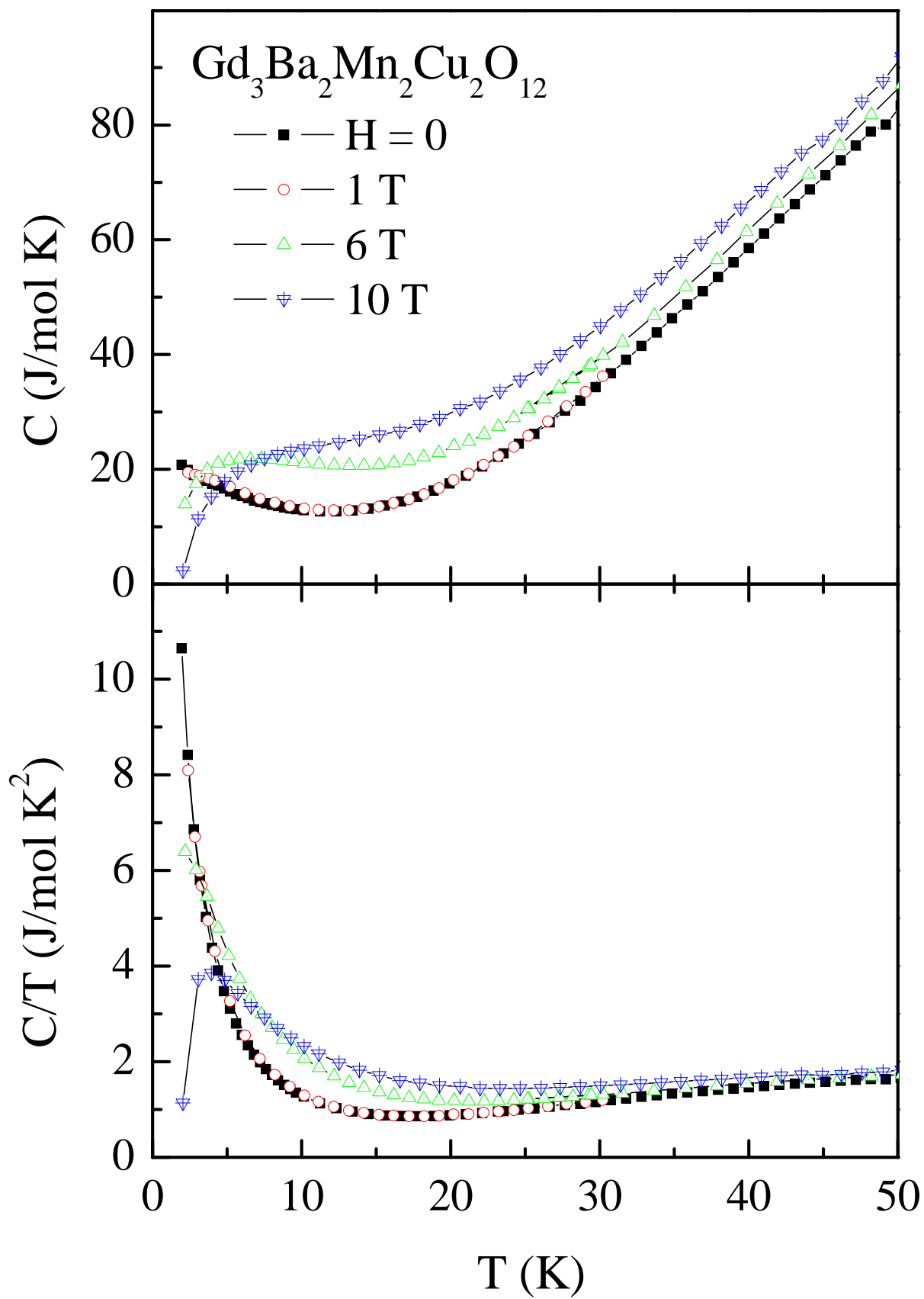

Fig. 6 Rayaprol et al

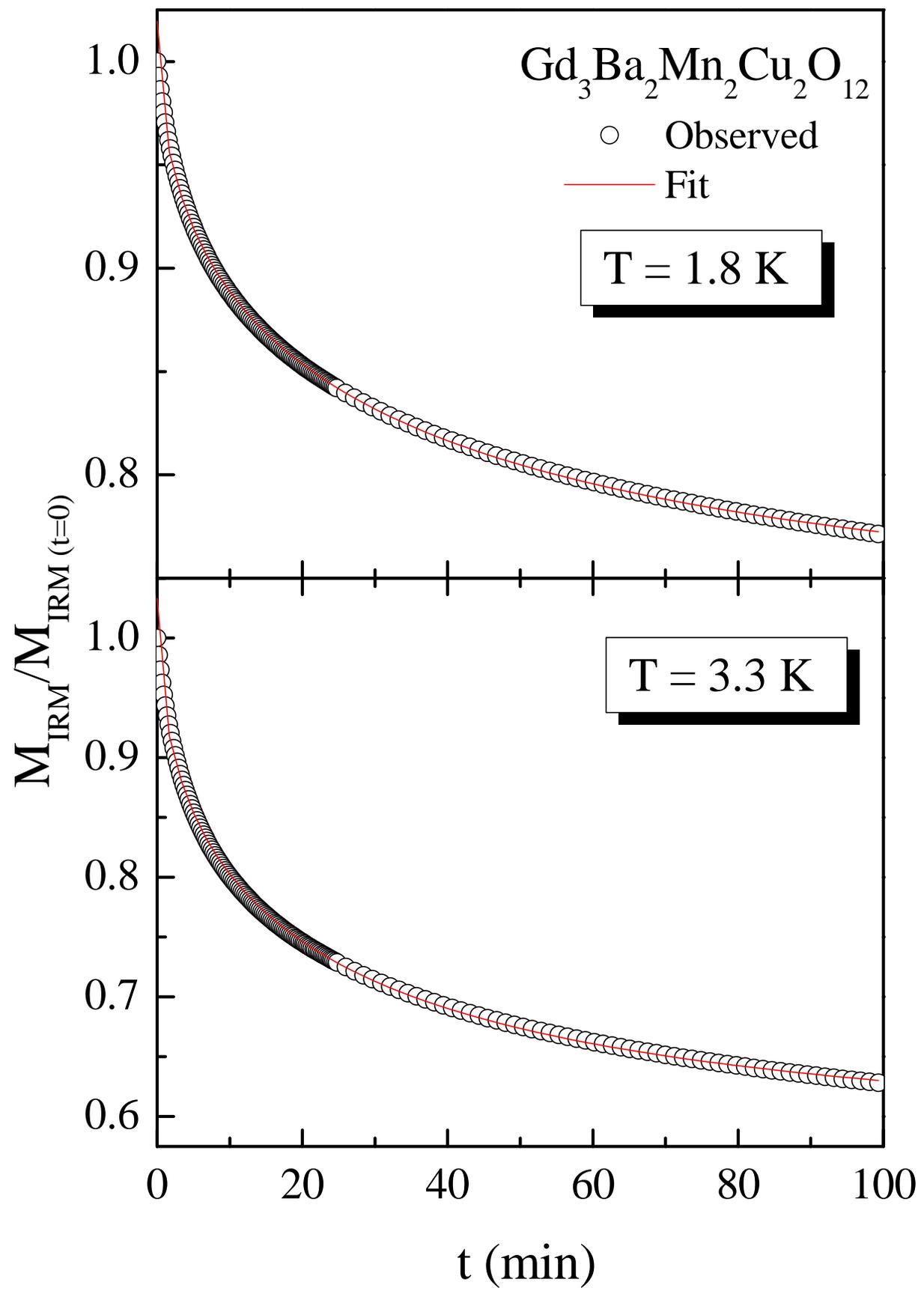

Fig 7. Rayaprol et al

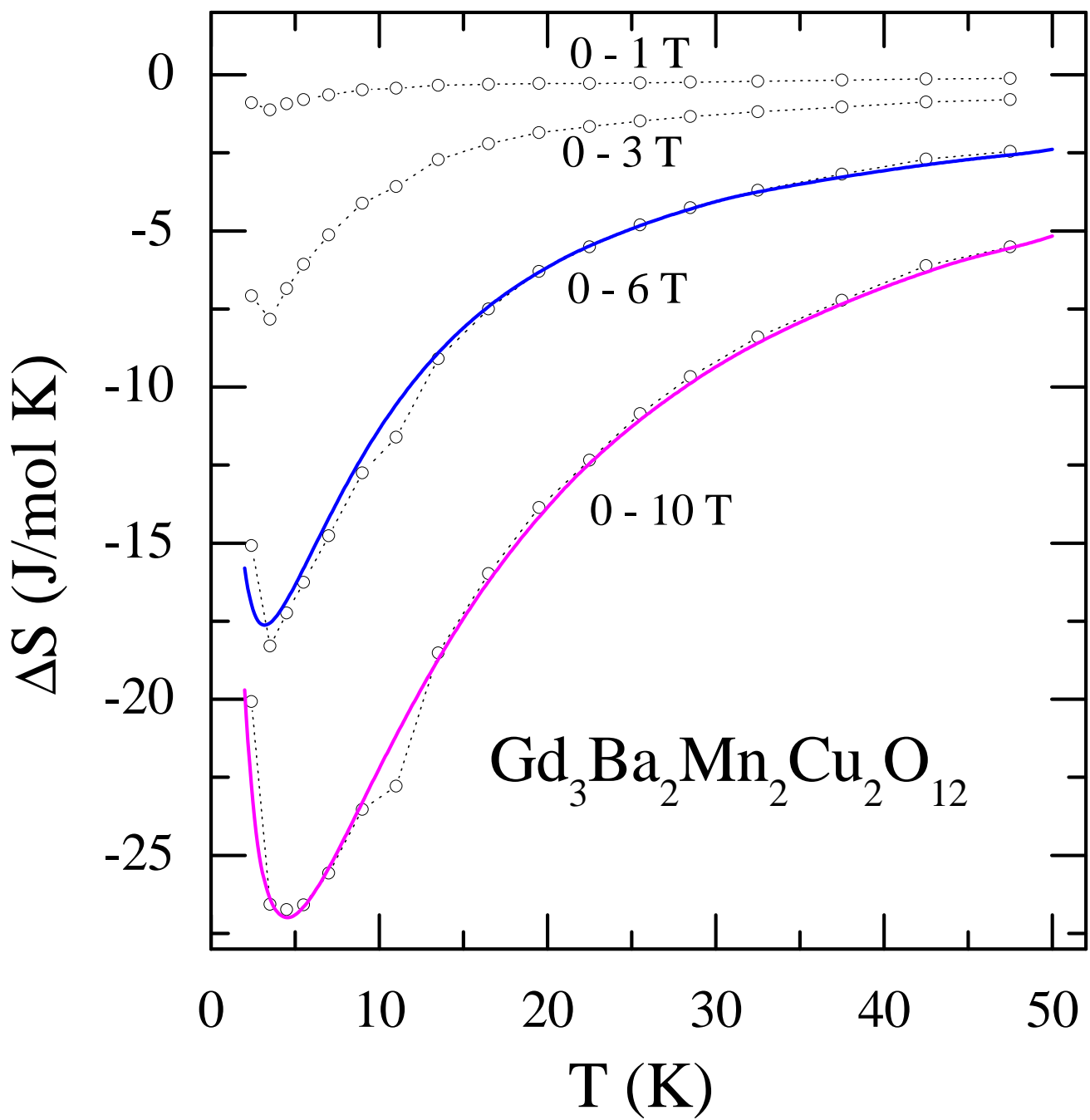

Fig. 8  Rayaprol et.al